\documentclass[conference]{IEEEtran}
\IEEEoverridecommandlockouts

\usepackage{url}
\usepackage{cite}
\usepackage{amsmath,amssymb,amsfonts}
\usepackage{algorithmic}
\usepackage{graphicx}
\usepackage{textcomp}
\usepackage{xcolor}
\usepackage{algorithm}

\title{LEED: A Highly Efficient and Scalable LLM-Empowered Expert Demonstrations Framework for Multi-Agent Reinforcement Learning}

\author{
\textit{Tianyang Duan}\textsuperscript{\rm 1},
    \textit{Zongyuan Zhang}\textsuperscript{\rm 1},
    \textit{Songxiao Guo}\textsuperscript{\rm 1},  \textit{Dong Huang}\textsuperscript{\rm 2}, \textit{Yuanye Zhao}\textsuperscript{\rm 3},
     \textit{Zheng Lin}\textsuperscript{\rm 4}, \\ \textit{Zihan Fang}\textsuperscript{\rm 5}, 
    \textit{Dianxin Luan}\textsuperscript{\rm 6}, \textit{Heming Cui}\textsuperscript{\rm 1}, \textit{Yong Cui}\textsuperscript{\rm 7}\\
    \textsuperscript{\rm 1} Department of Computer Science, The University of Hong Kong, Hong Kong, China.\\
    \textsuperscript{\rm 2} Institute of Data Science, National University of Singapore, Singapore.\\
    \textsuperscript{\rm 3} College of International Education, Hebei University of Economics and Business, China.\\
    \textsuperscript{\rm 4} Department of Electrical and Electronic Engineering, The University of Hong Kong, Hong Kong, China. \\
    \textsuperscript{\rm 5} Department of Computer Science, City University of Hong Kong, Hong Kong, China.\\
    \textsuperscript{\rm 6} Institute for Imaging, Data and Communications, University of Edinburgh, UK. \\
        \textsuperscript{\rm 7} Department of Computer Science and Technology, Tsinghua University, China. \\
}

\begin{document}

%
\maketitle
\begin{abstract}
Multi-agent reinforcement learning (MARL) holds substantial promise for intelligent decision-making in complex environments. However, it suffers from a coordination and scalability bottleneck as the number of agents increases. To address these issues, we propose the LLM-empowered expert demonstrations framework for multi-agent reinforcement learning (LEED). LEED consists of two components: a demonstration generation (DG) module and a policy optimization (PO) module. Specifically, the DG module leverages large language models to generate instructions for interacting with the environment, thereby producing high-quality demonstrations. The PO module adopts a decentralized training paradigm, where each agent utilizes the generated demonstrations to construct an expert policy loss, which is then integrated with its own policy loss. This enables each agent to effectively personalize and optimize its local policy based on both expert knowledge and individual experience. Experimental results show that LEED achieves superior sample efficiency, time efficiency, and robust scalability compared to state-of-the-art baselines.
\end{abstract}

\section{Introduction}
\label{sec:intro}

Multi-agent reinforcement learning (MARL) has emerged as a powerful paradigm for addressing complex decision-making problems in multi-agent systems, with broad applications such as the Internet of Things \cite{lin2024adaptsfl,hady2025multi,lin2024efficient,lin2025hierarchical}, robotic collaboration \cite{sun2025intra,orr2023multi,lin2023pushing,sun2025rrto}, and traffic signal control \cite{zhao2024survey,fang2024ic3m}. Existing MARL methods fall into two paradigms. Decentralized approaches \cite{yu2022surprising,de2020independent} train agents independently, emphasizing scalability and adaptability. In contrast, centralized training with decentralized execution (CTDE) \cite{rashid2020monotonic,ackermann2019reducing} uses centralized policy evaluation during training to promote coordination and improve overall team performance at execution time.

Despite recent advances in MARL, achieving scalable and reliable coordination remains a significant challenge \cite{zhang2021multi,zhang2025robust}. Although decentralized methods scale well with the number of agents, they restrict each agent to local observations and individual rewards, limiting accurate modeling of the policies of other agents, often producing policy conflicts, particularly in settings with global rewards. CTDE methods mitigate policy conflicts. However, as the number of agents increases, the joint state–action space in MARL grows exponentially, leading to extensive optimization costs and amplifying value-estimation errors when using function approximation. These errors can cause the homogenization of action values across agents, reducing the behavioral diversity and impairing the ability to find the optimal policy.

Large language models (LLMs)~\cite{lin2025hsplitlora,fang2025dynamic} have shown strong ability to abstract high-dimensional state–action spaces and address complex decision-making problems~\cite{chen2025towards,lin2024fedsn}. Recent LLM-driven advances, such as state representation extraction \cite{wang2024llm}, sub-task composition \cite{wang2023describe}, and reward design \cite{kwon2023reward} have significantly improved single-agent adaptability and generalization. However, how to integrate LLM-derived knowledge with policy optimization to achieve highly efficient and scalable MARL remains largely unexplored.

To this end, we propose the \underline{L}LM-\underline{e}mpowered \underline{e}xpert \underline{d}emonstrations framework (LEED), a decentralized MARL framework that integrates the extensive domain knowledge of LLMs with autonomous policy learning. LEED comprises two key components: a Demonstration Generation (DG) module and a Policy Optimization (PO) module. The DG module uses LLMs to generate environment-specific instructions, providing demonstrations about each agent’s task. These demonstrations are iteratively refined using environmental feedback to ensure relevance and efficacy. The PO module adopts a decentralized training paradigm, where each agent incorporates the generated demonstrations into its own learning process by constructing a mixed policy loss. This mixed loss enables agents to customize and optimize their local policies by striking an optimal balance between LLM-generated guidance and autonomous exploration. We evaluate LEED on two challenging, real-world multi-agent urban traffic planning tasks. Results show that LEED has superior training performance relative to state-of-the-art baselines.

\vspace{-5pt}
\section{Related Work}
\vspace{-5pt}
MARL has attracted significant attention for addressing complex cooperative and competitive tasks involving multiple agents \cite{zhang2021multi}. Decentralized methods, such as MAPPO, improve sample efficiency via shared policies \cite{yu2022surprising}. IPPO \cite{de2020independent} extends PPO \cite{schulman2017proximal} to the multi-agent setting by training independent policies for each agent without parameter sharing. Communication-enhanced approaches enable better coordination through information exchange \cite{muller2024clustercomm,lidard2022provably}. CTDE methods estimate the expected returns of joint actions through Q-functions and derive optimal joint policies accordingly \cite{qu2019exploiting,yuan2023multi,sunehag2017value}. QMIX achieves efficient joint Q-function estimation through monotonic value function factorization \cite{rashid2020monotonic}. Building on QMIX, HMDQN integrates hierarchical structure to mitigate sparse reward in multi-robot tasks \cite{bai2023smart}. HATRPO enhances stability in cooperative scenarios via advantage decomposition \cite{kuba2021trust}. MACPO incorporates safety constraints into MARL, ensuring agents consistently satisfy constraints during policy updates via trust region restrictions \cite{gu2021multi}. Deterministic policy methods output actions directly rather than probability distributions and improve policy robustness by leveraging global information for centralized training \cite{lowe2017multi,ackermann2019reducing}. 

\section{Methodology}

\subsection{Problem Formalization}
MARL is formulated as a decentralized partially observable Markov process (Dec-POMDP) \cite{oliehoek2016concise}, defined by the tuple $\left \langle \mathcal{A}, \mathcal{S}, \boldsymbol{\mathcal{U}}, P, \left \{ R^i \right \}_{i\in \mathcal{A}} ,\Omega ,\left \{ O^i \right \}_{i\in \mathcal{A}}, \gamma \right \rangle$. $\mathcal{A} =\left \{ 1, \dots, n \right \}$ denotes the set of $n$ agents, $\mathcal{S}$ denotes the state space, and $\boldsymbol{\mathcal{U}} = \prod_{i=1}^n \mathcal{U}^i$ denotes the joint action space. At each time step $t$, each agent $i$ receives a local observation $o_t^i$ from the observation space $\Omega$, according to the observation function $O^i(s_t)$. Agents select actions $u^i_t \sim \pi^i(\cdot \mid o^i_t)$, forming a joint action $\mathbf{u}_t = (u_t^1, \dots, u_t^n)$. The environment transitions to the next state $s_{t+1} \sim P(\cdot \mid s_t, \mathbf{u}_t)$ and provides each agent with a reward $r^i_t=R^i(s_t, \mathbf{u}_t, s_{t+1})$. The process continues, with each agent receiving an observation-action-reward tuple $(o^i_t,u^i_t,r^i_t)$, which is formalized as follows:
\small
\begin{equation}
    \boldsymbol{\tau }^a \sim \text{Env}(\boldsymbol{\pi  } \mid s_0,P),
\end{equation}
where $\boldsymbol{\tau}^a=\left \{ \tau^{a,i}\mid i\in \mathcal{A}  \right \} $ denotes the set of observation trajectories for all agents, $\boldsymbol{\pi}= {\textstyle \prod_{i=1}^{n}} \pi^i$ denotes the joint policy, $\tau^{a,i}=\left(o^i_0, u^i_0, r^i_0, o^i_1, u^i_1, r^i_1,\dots\right)$ denotes the observation trajectory for agent $i$, and $\text{Env}(\cdot)$ denotes the environment in which agents interacts. The objective of all agents is to learn a joint policy $\boldsymbol{\pi}$ to maximize the expected discounted return $\mathbb{E} \left [  {\textstyle \sum_{t=0}^{\infty }\textstyle \sum_{i=1}^{n }\gamma ^t R^i(s_t, \mathbf{u}_t, s_{t+1}) } \right ]$, where $\gamma \in [0, 1]$ denotes the discount factor.


\begin{figure}[!t]
  \centering
    \includegraphics[width=0.90\linewidth]{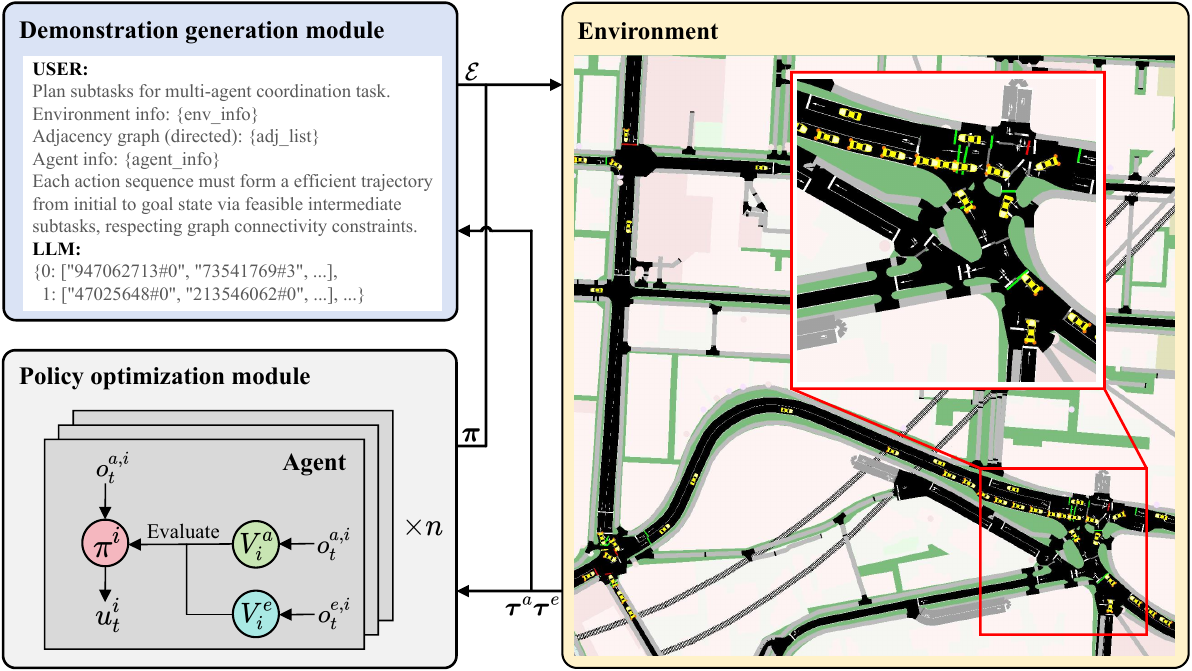}
      \vspace{-1.5ex}
    \caption{The workflow of LEED.}
    \vspace{-1ex}
    \label{fig:workflow}
\end{figure}

\subsection{Decentralized LLM-Empowered Multi-Agent Reinforcement Learning Framework}

As illustrated in Fig.~\ref{fig:workflow}, LEED framework consists of two components: the demonstration generation (DG) module and the policy optimization (PO) module. The DG module employs the LLM to generate an executable sequence set $\mathcal{E}$, which is executed in the environment to obtain expert demonstration trajectories $\boldsymbol{\tau}^e$. In the PO module, these trajectories, together with autonomously explored trajectories $\boldsymbol{\tau}^a$ collected by the joint policy $\boldsymbol{\pi}$ during environmental interaction, are used to optimize each agent's policy.

\subsubsection{Demonstration Generation Module}

The LLM generates an executable sequence set $\mathcal{E} = \left \{ \mathbf{e} ^i \mid  i \in  \mathcal{A}\right \}$ in the DG module, where each sequence $\mathbf{e} ^i=\left [ e^i_1,e^i_2,\dots  \right ] $ comprises a series of structured textual instructions. Each instruction $\boldsymbol{e}^i_k$ is formatted as \texttt{[Command, Parameter, AgentID]}, where \texttt{Command} specifies the action to be executed, \texttt{Parameter} provides associated arguments, and \texttt{AgentID} identifies the target agent. For example, in vehicle routing, \texttt{[MoveTo, Intersection\textbackslash$\_$5, Agent\textbackslash$\_$3]} directs Agent 3 to navigate to Intersection 5. Agents execute $\mathcal{E}$ in the environment to obtain expert demonstration trajectories:
\small  
\begin{equation}
    \boldsymbol{\tau }^e  \sim \mathrm{Env}(\mathcal{E} \mid s_0,P),
\end{equation}
where $\boldsymbol{\tau}^e =  \left \{ \tau^{e,i} \mid i \in \mathcal{A}  \right \} $ denotes expert demonstration trajectories for all agents, and each $\tau^{e,i}$ has the same observation-action-reward sequence format as $\tau^{a,i}$. It is worth noting that superscripts $a$ and $e$ indicate the source of observation trajectories: $a$ refers to those obtained from agent-environment interactions, while $e$ denotes those collected by executing the sequence set $\mathcal{E}$ in the environment.

\subsubsection{Policy Optimization Module}

In the PO module, we extend PPO \cite{schulman2017proximal} to multi-agent scenarios, where the training and execution of each agent are decentralized to ensure scalability. Specifically, we define value functions for agent and expert trajectories for agent $i$:
\small
\begin{equation}
V^x_i(o^{x,i}_t) = \mathbb{E}_{\tau^{x,i}} \left[ \sum_{k=t}^{\infty } \gamma^k r^{x,i}_k \right],
\end{equation}
where $x \in \left \{ a, e \right \} $, and $o^{x,i}_t,r^{x,i}_k \in \tau^{x,i}$. Subsequently, we compute the finite-horizon bootstrapped return for both agent and expert trajectories:
\small
\begin{equation}
    \label{R}
    R^{x,i}_t = \sum_{k=0}^{T^x-t-1} \gamma ^k r^{x,i}_{t+k}+\gamma^{T^x-t} V^x_i\left ( o^{x,i}_{T^x}  \right )
\end{equation}
where $r^{x,i}_{t+k}, o^{x,i}_{T^x} \in \tau^{x,i}$, and $T^x$ denotes the length of the trajectory $\tau^{x,i}$. The agent value function $V^a_i$ and the expert value function $V^e_i$ are optimized by minimizing the mean squared error:
\small
\begin{equation}
    \label{Vloss}
    \mathcal{L}_{\mathrm{value}}(V^x_i) = \mathbb{E}_{\tau^{x,i}}  \left[ \left(V^x_i(o^{x,i}_t) - R^{x,i}_t \right)^2\right].
\end{equation}
The agent advantage $A^{a,i}_t$ and the expert advantage $A^{e,i}_t$ for trajectories $\tau^{a,i}$ and $\tau^{e,i}$ are computed using the corresponding value functions and returns:
\small
\begin{equation}
    \label{A}
    A^{x,i}_t = R^{x,i}_t - V^x_i(o^{x,i}_t).
\end{equation}

\begin{algorithm}[t]
\small
\caption{LEED Training Procedure} 
\label{alg:leed}
\begin{algorithmic}[1]
\REQUIRE LLM $\Pi$, prompt $\mathrm{d}$, sampling interval $q$.
\STATE \textbf{Initialize} $V^a_i$, $V^e_i$, and $\pi^i$ for each agent.
\FOR{each episode}
\STATE Sample $\boldsymbol{\tau }^a \sim \mathrm{Env}(\boldsymbol{\pi  } \mid s_0,P)$
\IF{$i \bmod q = 0$}
\STATE Sample $\mathcal{E}\sim \Pi \left ( \mathrm{d} \right ) $
\STATE Sample $\boldsymbol{\tau }^e  \sim \mathrm{Env}(\mathcal{E} \mid s_0,P)$
\ENDIF
\FOR{each agent}
\STATE Calculate $R^{x,i}_t$ and $A^{x,i}_t$ using Eq.~\ref{R} and~\ref{A}. 
\STATE Update $V^a_i$ and $V^e_i$ using Eq.~\ref{Vloss}.
\STATE Update $\pi^i$ using Eq.~\ref{Pittlloss}.
\STATE $\text{d}\gets \mathrm{d} \cup \left \{ \boldsymbol{\tau }^a,\boldsymbol{\tau }^e\right \} $
\ENDFOR
\ENDFOR
\end{algorithmic}
\end{algorithm}

We employ the clipped surrogate objective to formulate both the agent policy loss $\mathcal{L}^a$ and the expert policy loss $\mathcal{L}^e$ as follows:
\small
\begin{equation}
    \mathcal{L}^x (\pi^i) = \mathbb{E}_{\tau^{x,i}}\left[ \min(\omega ^{x,i}_t , \mathrm{clip}(\omega ^{x,i}_t, 1 - \epsilon , 1 + \epsilon  ))A^{x,i}_t\right],
\end{equation}
where $\omega^{x,i}_t = \pi^i(u^{x,i}_t|o^{x,i}_t) / \pi^{\mathrm{old},i}(u^{x,i}_t|o^{x,i}_t)$ denotes the importance sampling ratio, $\pi^{\mathrm{old},i}$ denotes the policy of agent $i$ from the previous iteration, and $\epsilon \in (0, 1)$ represents the clipping coefficient. We then define a mixed policy loss function $\mathcal{L}_\mathrm{mix}$ as follows:
\small
\begin{equation}
\label{Pimaxloss}
\mathcal{L}_\mathrm{mix} (\pi^i) = \alpha\mathcal{L}^a (\pi^i) + (1-\alpha) \mathcal{L}^e (\pi^i),
\end{equation}
where $\alpha = \exp\left(-k/K \cdot \mathrm{D}_\mathrm{DTW}(\tau^{a,i}, \tau^{e,i})\right)$ dynamically adjusts the weighting between the agent’s and expert’s policy losses, with $k$ and $K$ denoting the current and total training epochs, respectively, and $\mathrm{D}_\mathrm{DTW}(\tau^{a,i}, \tau^{e,i})$ denotes the dynamic time warping (DTW) distance, which measures the similarity between agent and expert trajectories via nonlinear temporal alignment. Early in training, a large DTW distance results in a smaller $\alpha$, placing greater weight on expert trajectories for effective imitation learning. As training progresses and trajectories align, $\alpha$ increases, gradually shifting the training focus toward the autonomous exploration.

\begin{figure}
    \centering
    \includegraphics[width=0.9\linewidth]{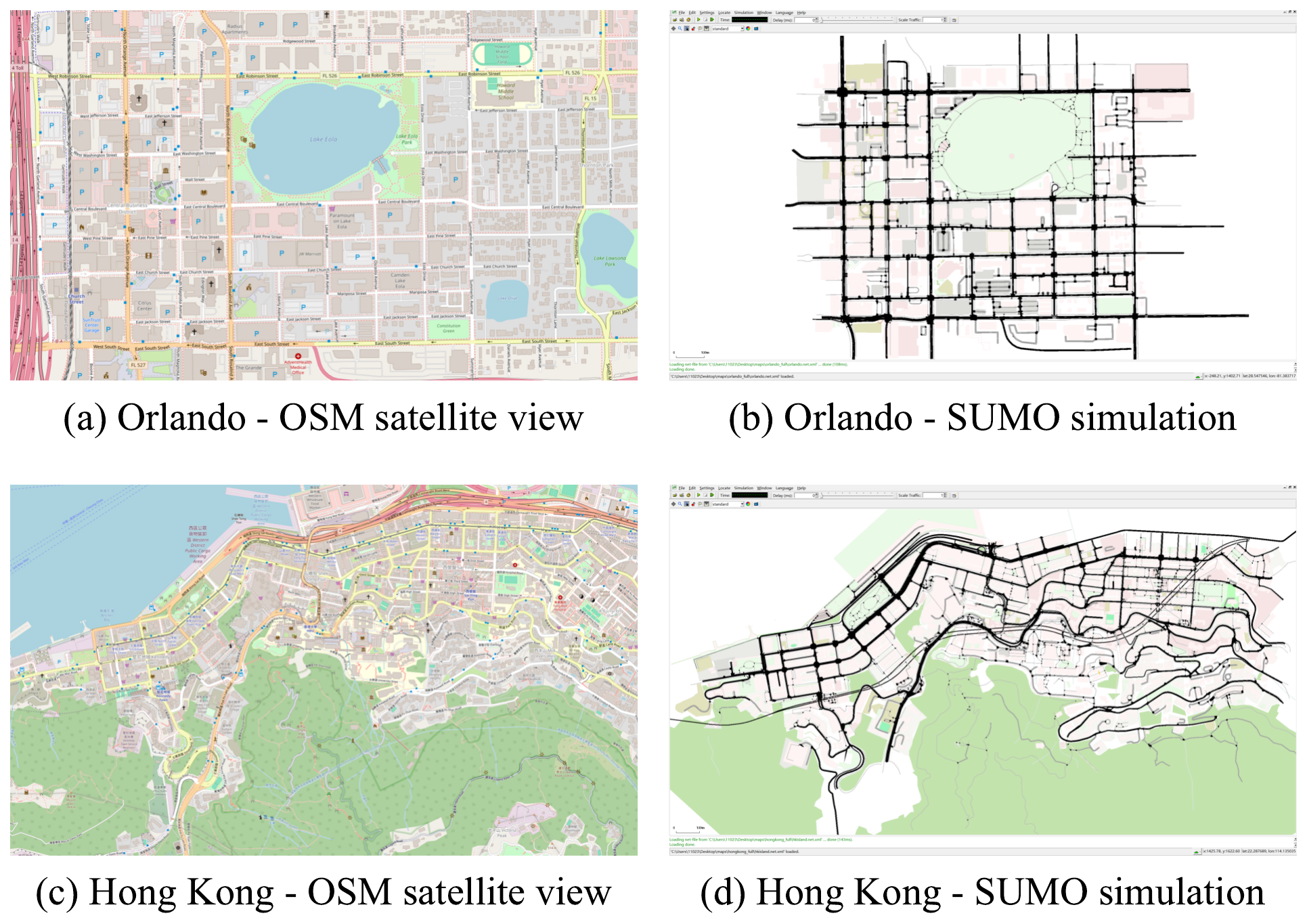}
    \caption{Experimental scenarios.}
    \label{fig:0_maps}
\end{figure}

\begin{table}[h]
\centering
\small 
\resizebox{0.4\textwidth}{!}{
\begin{tabular}{c|l}
\hline
\multicolumn{2}{c}{\textbf{Action Space (Discrete, shape: $m$)}} \\
\hline
\textbf{Index} & \textbf{Note} \\
\hline
$0..m-1$ & Select the $i$-th available edge \\
\hline
\multicolumn{2}{c}{\textbf{Observation Space (Shape: $2 + 2m$)}} \\
\hline
\textbf{Index} & \textbf{Note} \\
\hline
$[0]$ & Current junction id \\
$[1]$ & Destination junction id \\
$[2 + 2i]$ & Score of the $i$-th available edge ($i=0..m-1$) \\
$[2 + 2i + 1]$ & \begin{tabular}[t]{@{}l@{}} End junction id of the $i$-th edge ($i=0..m-1$) \end{tabular} \\
\hline
\end{tabular}
}
\caption{Action space and observation space}
\label{tab:action_obs_space}
\end{table}

Finally, PO module introduces a maximum entropy regularization term to each agent's policy loss to promote exploration. Thus, the policy loss $\mathcal{L} (\pi^i)$ for each agent is:
\small
\begin{equation}
    \label{Pittlloss}
   \mathcal{L} (\pi^i)=\mathcal{L}_\mathrm{mix} (\pi^i) + \beta \mathbb{E}_{\tau^{a,i}}[ \mathcal{H}(\pi^i(\cdot|o^{a,i}_t)) ],
\end{equation}
where $o^{a,i}_t \in \tau^{a,i}$, and $\beta \in \left(0, 1\right]$ denotes the entropy regularization coefficient, and $\mathcal{H}(\pi^i(\cdot|o^{a,i}_t))$ denotes the entropy of the policy $\pi^i$. 

Algorithm~\ref{alg:leed} outlines the training procedure of LEED, including the periodic refinement of the prompt $\mathrm{d}$ using both agent and expert trajectories (line 12). To stabilize updates and reduce inference overhead, we employ a sampling interval $q$ that reuses the same expert trajectories across multiple episodes (lines 4–7).

\section{Evaluation}
\label{sec:evaluation}
\subsection{Experimental Setup}
\subsubsection{Environment}
We evaluate LEED across challenging multi-agent cooperative navigation scenarios in large-scale traffic environments based on real-world OpenStreetMap (OSM) data \cite{OpenStreetMap}, with simulations implemented via SUMO \cite{SUMO2018}. Experiments span two representative urban networks: \textit{(1) Orlando}, a canonical grid with uniform intersections (Figure~\ref{fig:0_maps}(a,b)), and \textit{(2) Hong Kong}, a complex, mountainous network with heterogeneous road geometries and non-standard intersections (Figure~\ref{fig:0_maps}(c,d)). The reward function includes a time penalty, a distance-based shaping term toward the destination, and a completion bonus upon arrival. Detailed environment specifications are summarized in Table~\ref{tab:action_obs_space}.

\subsubsection{Prompt Design}

We use GPT-3.5 to generate LLM-based expert demonstrations for LEED. We provide the LLM with task description, node coordinates, a directed graph adjacency list, and agent information (start/end points and departure times), establishing context for the traffic environment. The LLM is instructed to output parsable waypoint sequences for each agent as Python dictionary which is executable in the simulation. The refinement prompt is generated every $q=5$ epochs using trajectory data from both agent exploration $\tau^a$ and expert demonstrations $\tau^e$. DTW distances are computed to determine the mixing weight $\alpha$ in Eq.~\ref{Pimaxloss} and analyze the differences between these trajectories.

\subsubsection{Baselines and Hyperparameters}

We compare LEED with several state-of-the-art MARL baselines (IPPO \cite{de2020independent}, MAPPO \cite{yu2022surprising}, QMIX \cite{rashid2020monotonic}) across two scenarios. All experiments are conducted with 5 independent runs. Each run consists of 500 training epochs, with each epoch taking 200 steps, totaling $1 \times 10^5$ steps. We use 10 agents and a learning rate of $3 \times 10^{-4}$. Both policy and critic networks use a 2-layer architecture with 128 cells per layer.

\subsection{Experimental Results and Analysis}

\subsubsection{Sample Efficiency}

Figure~\ref{fig:1_traning_curve}(a,b) shows LEED consistently surpasses all baselines in both sample efficiency and final performance. In \textit{Orlando}, LEED achieves $\sim500+$ reward while the best-performing baseline achieves only $\sim400$.  In the more challenging \textit{Hong Kong}, LEED maintains $\sim1000+$ reward compared to $\sim750-800$ for others.

\subsubsection{Time Efficiency}

\begin{figure}
    \centering
    \includegraphics[width=1\linewidth]{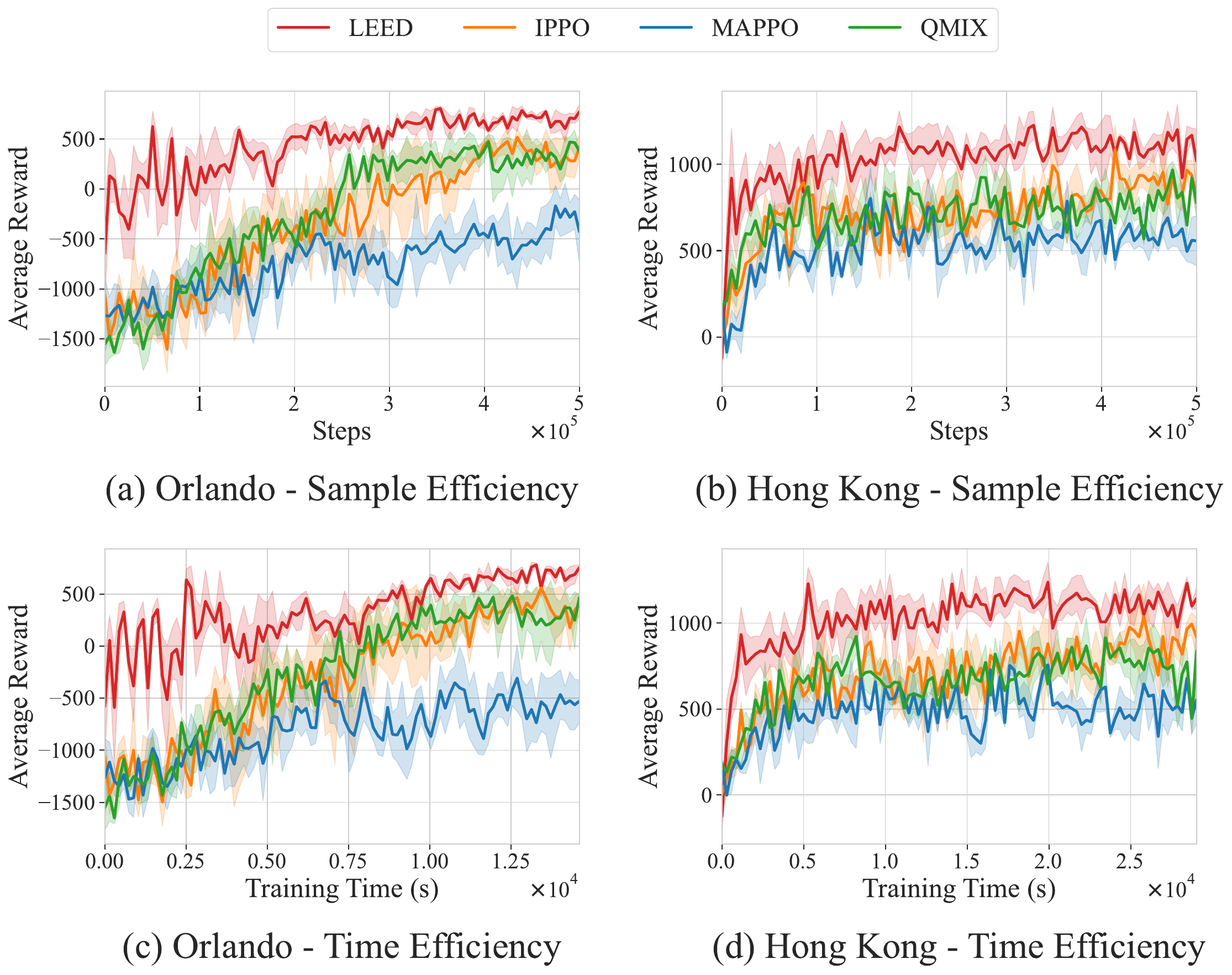}   
    \caption{Comparative results for sample and time efficiency.}
    \label{fig:1_traning_curve}
\end{figure}

Figure~\ref{fig:1_traning_curve}(c,d) shows that despite incorporating LLM inference during training, LEED remains competitive in terms of wall-clock training time. It achieves optimal performance efficiently in both scenarios.

\subsubsection{Scalability Analysis}
Figure \ref{fig:2_ablation}(a) presents a comparative performance in \textit{Orlando} under different agent population sizes (5, 10, 15, 20), reporting results over 20 test runs of the final model. LEED maintains better performance as agent numbers increase. While all methods show declining rewards when scaling from 5 to 20 agents, LEED experiences the smallest decrease, consistently achieving the highest average rewards. As the system scale increases, LEED can effectively alleviate policy conflicts among agents and have better scalability in large-scale systems.

\subsubsection{Ablation Study}
\begin{figure}
    \centering
    \includegraphics[width=1\linewidth]{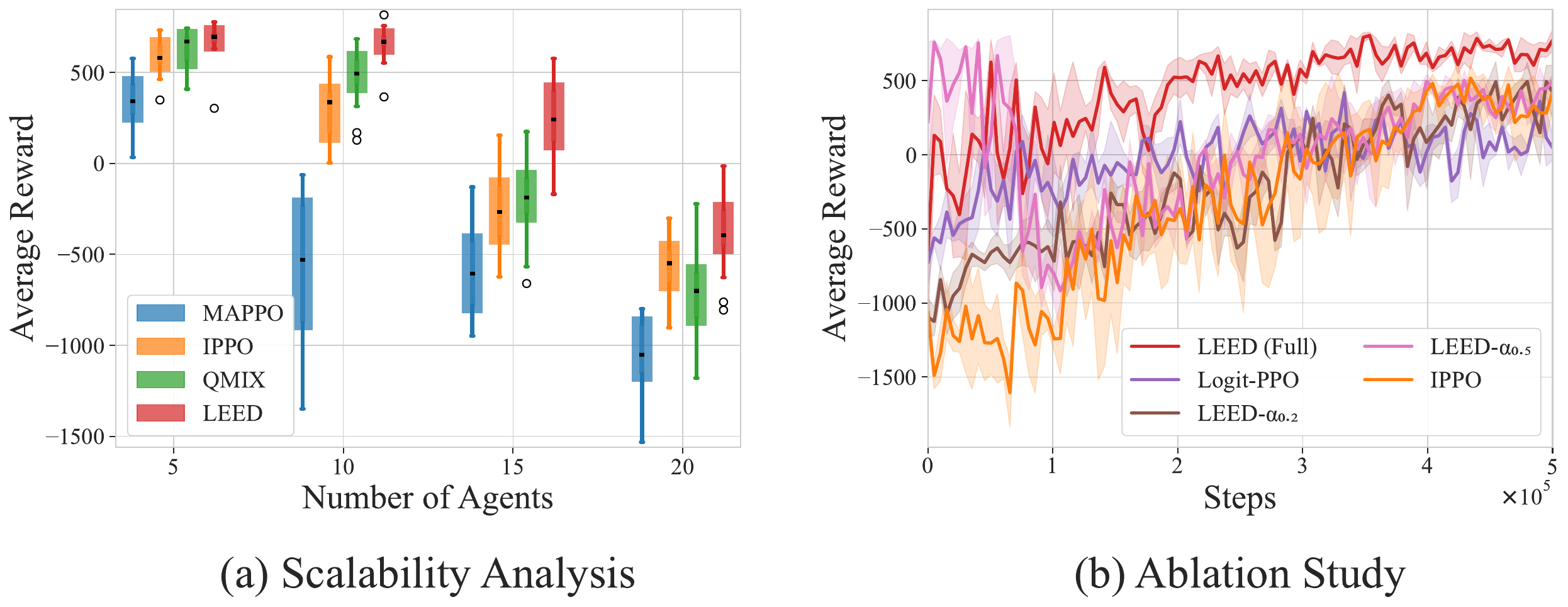}
    \caption{Scalability analysis and ablation study.}
    \label{fig:2_ablation}
\end{figure}

Figure \ref{fig:2_ablation}(b) presents our ablation study comparing LEED variants in \textit{Orlando}: \textit{(1) LEED (Full)}; \textit{(2) LEED-$\alpha_{0.2}$} and \textit{(3) LEED-$\alpha_{0.5}$} with fixed expert weights; \textit{(4) Logit-PPO}, which generate demonstrations by probabilistically sampling routes based on their shortest-path cost from a Logit-based model; and \textit{(5) IPPO}. Results show that LEED (Full) outperforms all variants. LEED-$\alpha_{0.5}$ exhibits strong early performance, but shows poor convergence due to insufficient agent exploration. LEED-$\alpha_{0.2}$ shows a slower learning speed. This suggests that a dynamic loss weighting mechanism yields superior performance compared to static weighting. Logit-PPO performs better than IPPO, yet is inferior to LEED. This can be attributed to the lower quality of samples generated by random methods and the absence of a feedback mechanism.

\subsubsection{Quality of LLM-Generated Demonstrations}
Table~\ref{tab:llm_quality} presents our evaluation of LLM-generated demonstrations across three refinement phases (initial, after 5 refinements, and after 10 refinements) in \textit{Orlando}. For each phase, 100 expert trajectories (10 prompts × 10 agents) are collected. The results show progressive quality improvements with increased refinements: token count expanded from 3680 to 7576 due to growing context, path validity rates improved from 74\% to 100\%, and rewards increased from 478.42 to 503.26. Additionally, the decreasing DTW distance indicates greater consistency between agent and expert behaviors as refinements accumulated.

\begin{table}[h]
\centering
\resizebox{0.45\textwidth}{!}{
\begin{tabular}{l|c|c|c}
\hline
\textbf{Metric} & \textbf{Initial} & \textbf{5 Refinements} & \textbf{10 Refinements} \\
\hline
Token & 3680 & 4645 & 7576 \\
\hline
Validity Rate (\%) & 74 & 82 & 100 \\
\hline
Reward & 478.42 & 495.05 & 503.26 \\
\hline
DTW Distance & 189.91 & 50.53 & 41.25 \\
\hline
\end{tabular}
}
\caption{Analysis of LLM-Generated Demonstrations.}
\label{tab:llm_quality}
\end{table}

\section{Conclusion}
We propose LEED, a scalable and decentralized multi-agent reinforcement learning framework that integrates LLM-generated expert demonstrations with autonomous agent exploration. By leveraging a hybrid loss function, LEED adaptively balances expert knowledge and individual agent experience, enabling efficient and robust policy learning for each agent. Experimental results demonstrate that LEED not only enhances sample and time efficiency but also achieves superior scalability compared to state-of-the-art baselines. As a potential future direction, we are looking forward to extending our method to improve the performance of various applications such as large language models~\cite{fang2024automated,lin2024splitlora}, multi-modal training~\cite{tang2024merit,fang2025dynamic}, distributed machine learning~\cite{zhang2024fedac,lin2025leo,hu2024accelerating,zhang2025lcfed}, and autonomous driving~\cite{lin2022tracking,fang2024ic3m,lin2022channel}.

\bibliographystyle{IEEEbib}
\bibliography{refs}

\end{document}